\documentclass[aps,pra,superscriptaddress,amssymb,showpacs]{revtex4}

\usepackage{graphicx}

\begin{document}

% Use the \preprint command to place your local institutional report
% number in the upper righthand corner of the title page in preprint mode.
% Multiple \preprint commands are allowed.
% Use the 'preprintnumbers' class option to override journal defaults
% to display numbers if necessary
%\preprint{}

%Title of paper
\title{Excitation of coherent polaritons in a two-dimensional atomic lattice}

% repeat the \author .. \affiliation  etc. as needed
% \email, \thanks, \homepage, \altaffiliation all apply to the current
% author. Explanatory text should go in the []'s, actual e-mail
% address or url should go in the {}'s for \email and \homepage.
% Please use the appropriate macro foreach each type of information

% \affiliation command applies to all authors since the last
% \affiliation command. The \affiliation command should follow the
% other information
% \affiliation can be followed by \email, \homepage, \thanks as well.

\author{I.O. Barinov}
\author{A.P. Alodjants}
\email[]{alodjants@vpti.vladimir.ru}
\author{S.M. Arakelian}
%\homepage[]{Your web page}
%\thanks{}
%\altaffiliation{}
\affiliation{Department of Physics and Applied Mathematics, Vladimir
State University,  Gorkogo str.  87,  Vladimir, 600000 Russia}

%Collaboration name if desired (requires use of superscriptaddress
%option in \documentclass). \noaffiliation is required (may also be
%used with the \author command).
%\collaboration can be followed by \email, \homepage, \thanks as well.
%\collaboration{}
%\noaffiliation

\begin{abstract}
We describe a new type of spatially periodic structure (lattice
models): a polaritonic crystal (PolC) formed by a two-dimensional
lattice of trapped two-level atoms interacting with quantised
electromagnetic field in a cavity (or in a one-dimensional array of
tunnelling-coupled microcavities), which allows polaritons to be
fully localised. Using a one-dimensional polaritonic crystal as an
example, we analyse conditions for quantum degeneracy of a
low-branch polariton gas and those for quantum optical information
recording and storage.
\end{abstract}

% insert suggested keywords - APS authors don't need to do this
%\keywords{}
%\maketitle must follow title, authors, abstract, \pacs, and \keywords
\maketitle
\section{INTRODUCTION}
This work develops the ideas that S.A. Akhmanov paid much attention
to in the last years of his life. It addresses the general
principles, laid by Akhmanov, behind the optical information
recording and processing using non-linear imaging in spatially
periodic or inhomogeneous dynamic structures excited by laser
radiation in nonlinear media of various kinds (see, e.g., [1]).

In recent years, great advances have been made in laser control of
macroscopic amounts of ultracold atoms [2]. The ability to produce
an array of macroscopic atomic Bose--Einstein condensates (BECs) by
cooling and trapping atoms in one- and two-dimensional optical
lattices enables research into various physical aspects of phase
transitions. Strong atom--photon coupling has recently been
demonstrated for BEC atoms in a cavity [3]. Basically, the
fabrication of spatially periodic atomic structures confined in an
optical cavity opens up novel opportunities for gaining insight into
critical phenomena in coupled atom--photon systems [4]. Recent
advances in nanotechnology and photonics have made it possible to
create such structures using one-dimensional (1D) arrays of coupled
microcavities, so-called coupled-resonator optical waveguides,
containing two- or three-level atoms [5--7]. A key role in
determining the behaviour of such systems is played by bright- and
dark-state polaritons--bosonic quasi-particles resulting from a
linear superposition of a photon and macroscopic (coherent)
excitation of a two-level atomic system.

In Ref. [8] Bose--Einstein condensation and the Kosterlitz--Thouless
phase transition for polaritons resulting from the interaction of a
quantised light field with an ensemble of two-level atoms in a
cavity is proposed. Note that the phase transition in question may
take place at sufficiently high (room) temperatures because of the
low-branch polariton effective mass [9]. Kasprzak et al. [10]
demonstrated a macroscopical population of ground state (in-plane
wave number $k_{\|}$) of a 2D gas of exciton-polaritons in
semiconductor nanostructures (Cd--Te) at 5 K. The superfluid
properties and Josephson dynamics of such polaritons were studied by
Alodjants et al. [11] and observed experimentally by Lai et al.
[12]. In addition, as shown by Alodjants et al. [13] certain
conditions enable optical cloning and quantum memory based on the
cavity polaritons in question.

In this paper, we discuss models of polaritonic crystal (PolC),
which can be produced using existing technologies and procedures for
laser control of atoms. A remarkable feature of such structures is
the possibility of complete polariton localisation, an analogue of
light localisation in photonic crystals in nonlinear optics. This
effect markedly reduces the group velocity of an optical wave packet
propagating through the medium. At the same time, polaritonic
crystals can be used to observe BECs of low-branch polaritons.

\section{Models of polaritonic crystals:
basic equations} Consider two models of PolC. In model 1, an
ensemble of ultracold two-level atoms is confined in a deep optical
lattice (Fig. 1a). This can be done by a number of experimental
means, in particular by using a two-component (spinor) condensate of
atoms with internal levels $\left| {a} \right\rangle$ and $\left|
{b} \right\rangle$ [14, 15]. In this case, a 2D periodic structure
of elliptical (needle-like) atomic condensates can be produced using
interference of two standing waves (not shown in Fig. 1a) along the
$x$ and $y$ axes, respectively. The atomic ensembles will then
interact with quantised light field in the cavity along the $z$ axis
in the strong coupling regime (see Eq. (2) below).

\begin{figure}
\includegraphics{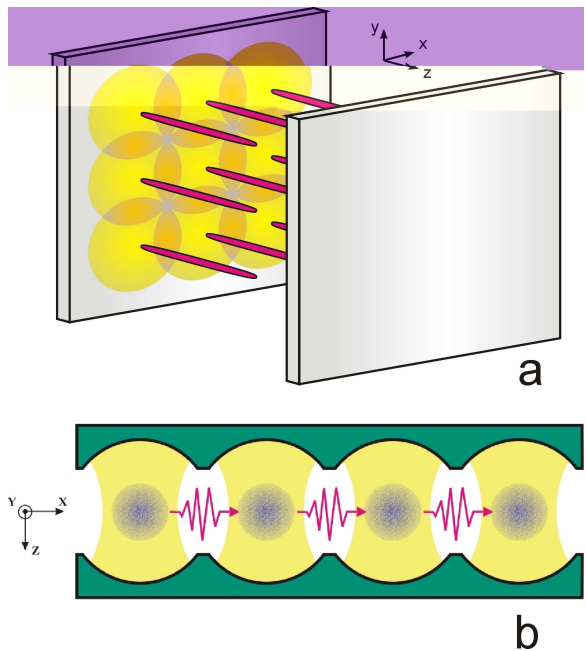}% Here is how to import EPS art
\caption{\label{fig:fig1} Schematic representation of the two models
for polaritonic crystals: (a) trapped ensembles of ultracold atoms
in a cavity interact with a quantised light field, whose
distribution is represented by grey circles; (b) polaritonic crystal
formed by an array of microcavities containing macroscopic ensembles
of two-level atoms.}
\end{figure}

Model 2 considers a lattice of $M$ tunnelling-coupled microcavities
in the $xy$ plane (Fig. 1b). Each cavity contains two-level atoms
interacting with the electromagnetic field along the $z$ axis. The
polaritonic crystals differ fundamentally from coupled-resonator
optical waveguides (see, e.g., [6]) in that they offer the
possibility of photon tunnelling in the $xy$ plane, normal to the
main axis of the cavities.

It can be shown that, in the limit of so-called strong coupling
between neighboring cells (cavities in model 2) containing atoms,
the two models schematically illustrated in Fig. 1 give the same
physical results. In this approximation, an atomic system can be
described as a system of bosons (bosonic modes) that evolve only in
time: its spatial degrees of freedom are frozen. This approximation
is valid when the number of atoms $N$ in each cell is relatively
small ($N\le 10^{4}$) [14]. The height of the potential barrier
between the cells of the optical lattice far exceeds the chemical
potential of each atomic ensemble [16]. Otherwise, it is necessary
to take into account the spatial configuration of the atomic system,
which leads to the formation of spatially localised atomic
structures [17].

Here, we restrict ourselves to the strong coupling approximation,
completely neglecting interatomic interactions and taking the atomic
ensembles in the cells to be an \textit{ideal gas}. It will become
clear from the analysis below that the polariton model for
atom-photon coupling is then quite correct.

To be more specific, we consider model 1 of polaritonic crystal. The
strong atom-photon coupling condition is thought to be fulfilled:
the coupling parameter $\kappa$ in each cell of the lattice is
substantially greater than the inverse of the coherence time,
$\tau_{coh}$, of the atom--photon system:
\begin{equation}
\kappa\gg \frac{1}{\tau_{coh}}.
\end{equation}

The total Hamiltonian of the system is
\begin{equation}
H=H_{at}+H_{int}+H_{ph},
\end{equation}
where
\begin{equation}
H_{at}=\sum\limits_{j=a,b}\int{\Phi_{j}^{+}\left(-\frac{\hbar^2\Delta}{2M_{at}}+V_{ext}^{(j)}\right)\Phi_{j}d^3\vec{r} },\\
\end{equation}
\begin{equation}
H_{ph}=\hbar \kappa\int{\left(\Psi^+\Phi_{a}^+\Phi_{b}+\Phi_{b}^+\Phi_{a}\Psi\right)d^3\vec{r} },\\
\end{equation}
\begin{equation}
H_{int}=\int{\Psi^{+}\left(-\frac{\hbar^2\Delta_{\bot}}{2M_{ph}}+V_{ph}\right)\Psi
d^2\vec{r} }.
\end{equation}
Here, $H_{at}$ represents the ensemble of noninteracting two-level
atoms in the trap; $H_{int}$ represents the atom ë photon
interaction in the cavity in the rotating wave approximation;
$H_{ph}$ represents the light field in the cavity in the paraxial
approximation; $\Phi_{j}(\Phi_{j}^+)$ are the boson annihilation
(creation) operators for the levels $j=a$ and $b$; $M_{at}$ is the
mass of a free atom; $\Delta$ is the Laplace operator;
$V_{ext}^{(j)}$ is the total atom trapping potential, which
comprises the harmonic potential of the magneto-optical trap and the
optical lattice potential along the $x$ and $y$ axes [18];
$\Psi(\Psi^+)$ is the annihilation (creation) operator for a field
propagating along the $z$ axis of the cavity, $M_{ph}=\hbar k_z/c$
is the photon effective mass in the cavity; $k_z$ is the $z$-axis
projection of the optical field wave vector; $V_{ph}$ is the photon
trapping potential in the atom-photon coupling region, which can be
created by special gradient-index lenses or fibres [8]; and
$\Delta_{\bot}=\partial^2/\partial x^2+\partial^2/\partial y^2$.

For an array of $M$ cells in an optical lattice, the $\Phi_{j}$ and
$\Psi$ operators can be represented in the form
\begin{equation}
\Phi_a=\sum\limits_{m=1}^{M}a_m(t)\varphi_{m}^{a}(\vec{r}),\\
\end{equation}
\begin{equation}
\Phi_b=\sum\limits_{m=1}^{M}b_m(t)\varphi_{m}^{b}(\vec{r}),\\
\end{equation}
\begin{equation}
\Psi=\sum\limits_{m=1}^{M}\psi_m(t)\xi_{m}(\vec{r}),\\
\end{equation}
where $\varphi_{m}^{a,b}(\vec{r})$ and $\xi_{m}(\vec{r})$ are the
real-valued Wannier functions describing the spatial distributions
of the atoms and field, respectively, in the $m$th cell
$(m=1,...,M)$. In the limit of strong coupling between neighbouring
cells (so-called tight-binding approximation), the
$\varphi_{m}^{a,b}$ functions satisfy the relations [16, 17]
\begin{equation}
\int|\varphi_m^{a,b}|^2 d^3\vec{r}=1,\quad
\int\varphi_m^{a,b}\varphi_{m+1}^{a,b} d^3\vec{r}\simeq 0.
\end{equation}
Analogous relations are valid for the $\xi_{m}(\vec{r})$ functions.
The $a_{m}(t)$ and $b_{m}(t)$ operators characterise the dynamic
behaviour of the two components (two modes) of the atomic ensemble
at the lower and upper levels, respectively, and the $\psi_{m}(t)$
operator describes the time evolution of the cavity field in the
$m$th cell of the lattice.

Substituting (6)-(8) into (3)-(5) we obtain
\begin{equation}
H_{at}=\hbar\sum\limits_{m=1}^{M}[\omega^{a}_{m at}a^{+}_ma_m+\omega^{b}_{m at}b^{+}_m b_m-\frac{\beta_a}{2}(a_m^+a_{m-1}+a_m^+a_{m+1}+h.c.)-\frac{\beta_b}{2}(b_m^+b_{m-1}+h.c.)],\\
\end{equation}
\begin{equation}
H_{int}=\hbar\sum\limits_{m=1}^{M}g_m(\psi_m^+a_m^+b_m+b_m^+a_m\psi_m),\\
\end{equation}
\begin{equation}
H_{ph}=\hbar\sum\limits_{m=1}^{M}[\omega_{m
ph}\psi_m^+\psi_m-\frac{\alpha}{2}(\psi_m^+\psi_{m-1}+\psi_m^+\psi_{m+1}+h.c.)],
\end{equation}
where the coupling coefficients $\beta_{a,b}$ and $\alpha$
characterise atom (photon) tunnelling between neighbouring cells and
are determined by the overlap integrals of the
$\varphi_m^{a,b}(\vec{r})$ and $\xi_m(\vec{r})$ functions with their
derivatives, respectively. The quantities $\omega_{m at}^{a,b}$ and
$\omega_{m ph}$ are defined in an analogous way [16, 17]. We take
all the atom-photon coupling coefficients to be the same in all the
cells: $g=g_1=g_2=...=g_M$.

Let us turn to the momentum ($\vec{k}$-space) representation. Given
that polaritonic crystals have a periodic structure, the $\psi_m$,
$a_m$ and $b_m$ operators can be represented in the form
\begin{equation}
a_m=\frac{1}{\sqrt{M}}\sum\limits_{\vec{k}} a_{\vec{k}}
exp(i\vec{k}\vec{n}),\quad
b_m=\frac{1}{\sqrt{M}}\sum\limits_{\vec{k}} b_{\vec{k}}
exp(i\vec{k}\vec{n}),\quad
\psi_m=\frac{1}{\sqrt{M}}\sum\limits_{\vec{k}}
\psi_{\vec{k}}exp(i\vec{k}\vec{n}),
\end{equation}
where $\vec{n}$ is a lattice vector.

For simplicity, we examine a 1D polaritonic crystal structure below,
for which $\vec{k}\vec{n}=mk_xl$, where $k_x$ is the $x$- axis
projection of the optical field wave vector and $l$ is the lattice
constant. Substituting (13) into (10)-(12), we obtain the
$\vec{k}$-space Hamiltonian \label{eq19}
\begin{equation}
H=\hbar\sum\limits_{\vec{k}}\left(\omega_{ph}(k)\psi_{\vec{k}}^{+}\psi_{\vec{k}}+\frac{1}{2}\omega_{at}(k)\left(b_{\vec{k}}^{+}b_{\vec{k}}-a_{\vec{k}}^{+}a_{\vec{k}}\right)+\frac{g}{\sqrt{M}}\sum\limits_{\vec{q}}\left(\psi_{\vec{k}}^{+}a_{\vec{q}}^{+}b_{\vec{k}+\vec{q}}+b_{\vec{k}+\vec{q}}^{+}a_{\vec{q}}\psi_{\vec{k}}\right)\right).
\end{equation}
Here $\omega_{ph}(k)$ and $\omega_{at}(k)$ determine the dispersion
relations for the photonic and atomic systems of the polaritonic
crystal, respectively, and are given by

\begin{equation}
\omega_{ph}(k)=\omega_{m ph}-2\alpha cos(kl),\quad
\omega_{at}(k)=\omega_{m at}^{b}-\omega_{m at}^{a}-2\beta cos(kl),
\end{equation}
where $\beta=\beta_b-\beta_a$ is the effective coupling coefficient
of the atomic lattice.
\section{Quantum degeneracy of a 1D polariton gas}
In the strong coupling regime, expression (14) is a many particle
Hamiltonian in the momentum representation, which describes a 1D
periodic structure and can be analysed in terms of dark- and
bright-state polaritons. We will consider it in the low atomic
excitation density limit, where all the atoms predominantly occupy
the lower level $|a\rangle$ [11]. The boson annihilation
($\phi_{\vec{k}}$) and creation ($\phi_{\vec{k}}^+$) operators for
collective excitations in a two-level atomic system can then be
defined in the $\vec{k}$-space representation:
\begin{equation}
\phi_{\vec{k}}=\sum_{\vec{q}}\frac{a_{\vec{q}}^{+}b_{\vec{k}+\vec{q}}}{\sqrt{MN}},\qquad
\phi_{\vec{k}}^+=\sum_{\vec{q}}\frac{b_{\vec{k}+\vec{q}}^{+}a_{\vec{q}}}{\sqrt{MN}}.
\end{equation}
Using (16), Hamiltonian (14) of the system can be represented in a
more convenient form:
\begin{equation}
H=\hbar\sum\limits_{\vec{k}}\left(\omega_{ph}(k)\psi_{\vec{k}}^{+}\psi_{\vec{k}}+\omega_{at}(k)\phi_{\vec{k}}^{+}\phi_{\vec{k}}+g\left(\psi_{\vec{k}}^{+}\phi_{\vec{k}}+\phi_{\vec{k}}^{+}\psi_{\vec{k}}\right)\right),
\end{equation}
where we denote again g instead of $g\sqrt{N}$ . Hamiltonian (17)
can be diagonalised using the Bogoliubov transformations
\begin{equation}
\Xi_{1,\vec{k}}=\mu_1\psi_{\vec{k}}-\mu_2\phi_{\vec{k}},\qquad
\Xi_{2,\vec{k}}=\mu_1\phi_{\vec{k}}+\mu_2\psi_{\vec{k}},
\end{equation}
where
\begin{equation}
\mu_{1,2}^2=\frac{1}{2}\left[1\mp
\frac{\delta\omega}{(\delta\omega^2+4g^2)^{1/2}}\right]
\end{equation}
are Hopfield coefficients satisfying the normalisation condition
$\mu_1^2+\mu_2^2=1$;
$\delta\omega=\omega_{at}(k)-\omega_{ph}(k)=\Delta-2(\beta-\alpha)\times\cos(kl)$
is the frequency detuning, dependent on quasimomentum $k$; and
$\Delta\equiv\omega_{m at}^b-\omega_{m at}^a-\omega_{m ph}$ is the
detuning at $kl=\pi/2+\pi p$ with $p=0,1,...$.

The $\Xi_{1\vec{k}}$ and $\Xi_{2\vec{k}}$ operators in (18)
represent two types of elementary excitations in an atomic system ë
upper and low branch polaritons, with characteristic frequencies
$\Omega_{1,2}(k)$, that determine the dispersion relations and band
structure of PolC. The frequencies are given by the expression
\begin{equation}
\Omega_{1,2}(k)=\frac{1}{2}\left[\omega_{at}(k)+\omega_{ph}(k)\pm\sqrt{\delta\omega^2+4g^2}\right],
\end{equation}

Using (20), one can find the mass of the upper (subscript 1) and low
(subscript 2) branch polaritons:
\begin{equation}
m_{1,2}=\frac{2m_{at}m_{ph}\sqrt{\tilde{\Delta}^2+4g^2}}{(m_{at}+m_{ph})\sqrt{\tilde{\Delta}^2+4g^2}\mp(m_{at}-m_{ph})\tilde{\Delta}}
\end{equation}
where $\tilde{\Delta}=\Delta-2(\beta-\alpha)$ is an effective
detuning, which includes the characteristic frequencies $\alpha$ and
$\beta$, $m_{at}=\hbar/(2\beta l^2)$ and $m_{ph}=\hbar/(2\alpha
l^2)$ are the effective masses of the atoms and photons in the
lattice, respectively.

Figure 2 shows the upper and low branch polariton dispersion,
$\Omega_{1,2}(k)$, in the first Brillouin zone. The band gap is
determined by the Rabi splitting: $(\delta\omega^2+4g^2)^{1/2}$. In
the center of the zone (near $k=0$), both dispersion branches are
parabolic. At resonance ($\delta\omega=0$), the splitting is
governed by the atom--photon coupling coefficient $2g$ (see also the
inset in Fig. 2).

The minimum at $k=0$ in the lower polariton branch in Fig. 2 is of
fundamental importance. At small magnitudes of the quasi-momentum,
$kl\ll1$, we have from (20)
\begin{equation}
\Omega_{2}(k)\approx\frac{\hbar k^2}{2m_{2}},
\end{equation}
which corresponds to the dispersion law of free particles
(polaritons) at the minimum in $\Omega_{2}(k)$ in Fig. 2. The
statistical properties of the polariton gas are then governed by its
dimensionality (see, e.g., [19]). In particular, in the case of
resonance coupling the low branch polariton mass can be found from
(21):
\begin{equation}
m_2=\frac{2m_{ph}}{1+m_{ph}/m_{at}}.
\end{equation}

\begin{figure}
\includegraphics{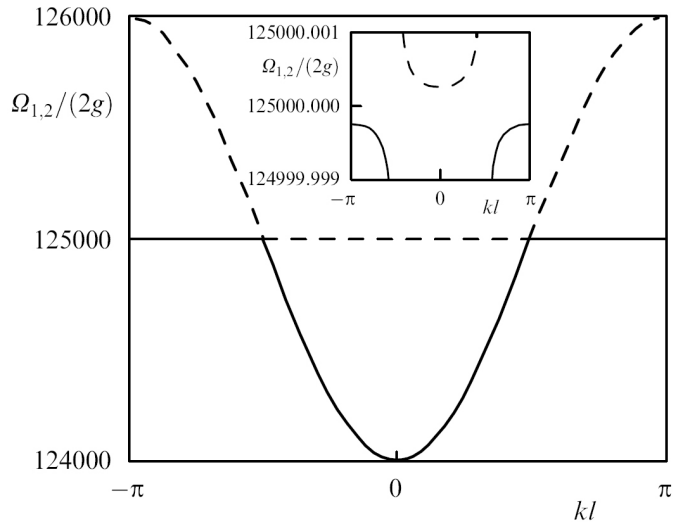}% Here is how to import EPS art
\caption{\label{fig:fig2} Polariton dispersion branches $\Omega_1$
(dashed curve) and $\Omega_2$ (solid curve) as functions of reduced
quasi-momentum (Bloch vector) in the first Brillouin zone. The
characteristic frequency of the atomic transition is $(\omega_{m
at}^b-\omega_{m at}^a)/2\pi=500THz$, $\Delta=0$, the atom--photon
coupling frequency is $g(2\pi)^{-1}=2GHz$, the photon (atom)
effective mass in the lattice is $m_{ph}=5\times 10^{-36}kg$
($m_{at}= 38.5\times 10^{-27}kg$), and the lattice constant is
$l=2.24\mu m$. The inset shows the frequency range of Rabi
splitting.}
\end{figure}
For $m_{ph}/m_{at}\ll 1$ ($\alpha\gg\beta$), the polariton mass is
sufficiently small. For example, in the case of interaction of
two-level sodium atoms with an electromagnetic field in a cavity at
a level separation wavelength of 589 nm, the photon effective mass
in the cavity is $m_{ph}\simeq 0.5\times 10^{-35}kg$ [12]. The
polariton mass estimated by Eq. (23) is then $m_2\simeq 10^{-35}kg$.
Therefore, the quantum degeneracy temperature of a 1D polariton gas,
$T_d=2\pi\hbar^2n_1^2/(m_2k_B)$, may be rather high ($\sim300K$ at a
polariton density $n_1\simeq10^4cm^{-1}$). It should, however, be
kept in mind that the spatially periodic structure of the
polaritonic crystal in Fig. 1 has a long coherence time only at
sufficiently low temperatures, where a relatively large number of
atoms can be trapped. The polariton gas can be considered a highly
degenerate quantum system, meeting the condition $n_1\Lambda_T\gg1$,
where $\Lambda_T=[2\pi\hbar^2/(m_2k_BT)]^{1/2}$ is the de Broglie
wavelength. In this limit, the formation of coherent polaritons in a
polaritonic crystal is of interest for spatially distributed
recording and storage of quantum optical information [20, 21].

\section{Group velocity of polaritons}
Consider the group velocities
$v_{1,2}=\partial\Omega_{1,2}(k)/\partial k$ of polaritons in a
lattice. From (20) we obtain
\begin{equation}
v_{1,2}=\frac{\hbar\sin{(kl)}}{2lm_{ph}}\left[1+\frac{m_{ph}}{m_{at}}\mp\left(1-\frac{m_{ph}}{m_{at}}\right)\frac{\delta\omega}{\sqrt{\delta\omega^2+4g^2}}\right]
\end{equation}
It follows from (24) that the group velocities $v_{1,2}$ of
polaritons are low at small $\vec{k}$. In particular, in the
$kl\ll1$ limit the low branch polariton has a linear
$v_2(k):v_2\simeq\hbar k/m_2$. At the same time, $v_{1,2}=0$ at the
boundaries of the Brillouin zone, i.e., at $kl=p\pi, p=0,\pm1,...$.
In this case, the structure of the polaritonic crystal allows
polaritons to be fully localised within this zone.

The ability to reduce the group velocity of polaritons can be used
to observe `slow' light, which is of high current interest for
quantum optical information recording and storage. In our case, the
group velocity of the optical field can be varied by changing the
atom-field detuning $\Delta$ (or $\tilde{\Delta}$).

The possibility of controlling low branch polaritons with small
magnitudes of the quasi-momentum, k, is exemplified in Fig. 3.
Control is performed near the bottom of the well in Fig. 2, where
the parabolic dispersion law (22) is valid. In this limit,
polaritons can be assigned a wave function, $\Psi(x,t)$, which
carries quantum information and satisfies free particle
Schr\"{o}dinger equation:
\begin{equation}
\left(i\hbar\frac{\partial}{\partial
t}+\frac{\hbar^2}{2m_2}\frac{\partial^2}{\partial
x^2}\right)\Psi(x,t)=0.
\end{equation}

The solution to Eq. (25) is well known in quantum physics (see,
e.g., [13, 22]). Polaritons are a coherent wave packet that broadens
with time and propagates through the polaritonic crystal (Fig. 1b).
The characteristic wave packet broadening time,
$\tau_b=m_2f^2/\hbar$, depends on both the polariton mass, $m_2$,
and the $x$-axis width of the packet, $f$, at the initial instant,
that is, on the incident beam diameter.

Quantum optical information recording and storage using PolC is
expected to have a three-step physical algorithm based on the
ability to control the group velocity of a polariton wave packet in
the medium by varying the $\tilde{\Delta}$ detuning (Fig. 3a).

For quantum information to be recorded in step I, the
$\tilde{\Delta}$ detuning must fulfil the condition
$\tilde{\Delta}\gg 2|g|$. The corresponding time interval in Fig. 3a
is $0\leq\tau\leq0.25$. The low branch polariton is here
photon-like, i.e., $\Xi_{2\vec{k}}\simeq-\psi_{\vec{k}}$
($\mu_1\simeq0, \mu_2\simeq-1$), and its effective mass $m_2\simeq
m_{ph}$ [see Eqs. (18), (19) and (21)]. The wave packet then
propagates at a velocity given by
\begin{equation}
v_2=\frac{\hbar k}{m_{ph}}=2\alpha l^2k,
\end{equation}
and displaces along the $x$ axis from the point $x=0$ to $x=v_2t$
(Fig. 3b). For example, at a quasi-momentum $k=10^5m^{-1}$ the
estimated polariton group velocity is $v_2=2\times 10^6m/s$.

To map the optical information to coherent excitations of the medium
in step II, the detuning should be made negative:
$\tilde{\Delta}\ll-2|g|$. In this limit, the low branch polariton
becomes atom-like, so that $\Xi_{2 \vec{k}}\simeq\phi_{\vec{k}}$
($\mu_1\simeq1, \mu_2\simeq0$) (Fig. 3). At
$m_{ph}/m_{at}=\beta/\alpha\ll g^2/\Delta^2$, its group velocity is
[7]
\begin{equation}
v_2=\frac{\hbar kg^2}{m_{ph}\tilde{\Delta}^2}=\frac{2\alpha
l^2kg^2}{\tilde{\Delta}^2}.
\end{equation}
When the more stringent condition
$\frac{g^2}{\tilde{\Delta^2}}\ll\frac{m_{ph}}{m_{at}}\ll1$ is
fulfilled, the low-branch polariton group velocity is
\begin{equation}
v_2=\frac{\hbar k}{m_{at}}=2\beta l^2k,
\end{equation}
which corresponds to the velocity of the atoms in the lattice. In
particular, for sodium atoms with an effective mass
$m_{at}=38.5\times10^{-27}kg$ the group velocity of such polaritons
at the above wave vector is $v_2=2.6\times10^{-4}m/s$ [21].

In fact, Eq. (28) determines the lower limit of the velocity of an
optical wave packet in the structure of a polaritonic crystal at a
given quasi-momentum. In this limit, all the information carried by
a light beam is recorded and stored via atomic excitations. At an
incident beam diameter $f\simeq10^{-4}m$, the estimated
characteristic broadening time is $\tau_b=m_{at}f^2/\hbar\approx3.7
s$. This time scale (the interval $0.25\leq\tau\leq0.75$ in Fig. 3a)
determines the longest information storage time in a quantum gas of
two-level sodium atoms. More precisely, a necessary condition for
such information recording is $t_{stor}\ll\tau_b$ (where $t_{stor}$
is the information storage time in the atomic system), which implies
that the wave packet retains its shape during the whole period of
quantum state storage.

\begin{figure}
\includegraphics{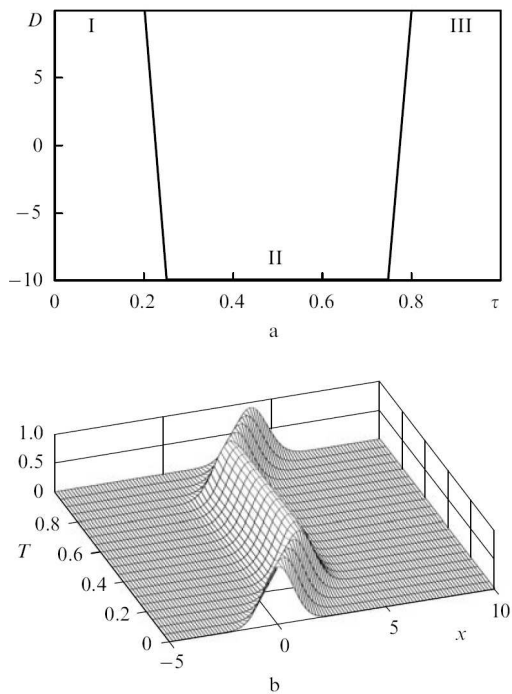}% Here is how to import EPS art
\caption{\label{fig:fig3} Propagation of a 1D polariton wave packet
through a polaritonic crystal: (a) $D=\tilde{\Delta}/(2|g|)$ as a
function of normalised time $\tau=\hbar t/(m_{ph}f^2)$ and (b) wave
packet envelope (probability density)
$S\equiv|\Psi(x,t)|^2/|\Psi(0,0)|^2$ as a function of $\tau$ and
normalised coordinate $X=x/f$ at $k_xf=10$; optical information (I)
recording, (II) storage and (III) reading (retrieving) steps.}
\end{figure}

To restore (read) optical information at the output of the medium
after time $t_{stor}$ (step III), the polaritons must be again made
photon-like by switching the $\tilde{\Delta}$ detuning in the
reverse direction. In Fig. 3a, the characteristic reverse switching
time of the wave packet, $t_{retr}$, is determined by the time
interval $0.75\leq\tau\leq0.8$. At the output ($\tau=1$), we again
have an optical wave packet, which is displaced along the $x$ axis
in a plane normal to the cavity axis.

In this work, we do not assess the quality (fidelity) of information
storage. Alodjants et al. [13] estimated the information storage
fidelity taking into account only changes in wave packet shape, but
this is generally insufficient. It is, in addition, necessary to
analyse transformations of the quantum state of the optical field
with consideration for the structure of the polaritonic crystal and
its decoherence (see, e.g., [23]). Analysis of this problem is of
interest on its own and is beyond the scope of this paper. We note
only that, under real experimental conditions, the information
recording, storage and reading time is limited by the decoherence
time of the polaritonic crystal. It is therefore quite reasonable to
use a condensate of atoms that have a sufficiently long macroscopic
coherence time--tens of microseconds according to experimental data
[24].

\section{Conclusions}
We examined a lattice model of coherent polaritons in a spatially
periodic structure a polaritonic crystal formed by a lattice of
ensembles of two-level atoms effectively interacting with quantised
electromagnetic field in a cavity (or in a 2D lattice of cavities)
in the strong coupling regime. Our results demonstrate that the
structure of PolC allows the low branch polariton to be fully
localised, which can be used, first, to achieve quantum degeneracy
of the polariton gas and, second, to substantially slow down group
velocity of light pulses in such media. The coherence properties of
an ensemble of polaritons were discussed from the viewpoint of
spatially distributed quantum recording, storage and retrieving of
information related to a propagating optical wave packet.

\section*{Acknowledgments}
This work was supported by the Russian Foundation for Basic Research
(Grant Nos 08-02-99011-r\_ofi, 08-02-99028-r\_ofi and
08-02-99013-r\_ofi) and the RF Ministry of Education and Science
through federal programmes.

\end{document}